\documentclass[oneside,twocolumn,epsfig]{revtex4}
\usepackage{graphicx}

\begin{document}
\title[] {Reaction-diffusion processes of three species on small-world networks}
\author{Kyungsik \surname{Kim$^{}$}}
\thanks{Electric mail: kskim@pknu.ac.kr;\\
Tel: +82-51-620-6354; Fax: +82-51-611-6357}
\author{K. H. \surname{Chang$^{a)}$}}
\author{Myung-Kul \surname{Yum$^{b)}$}}
\author{J. S. \surname{ Choi$^{}$}}
\author{T. \surname{Odagaki$^{c)}$}}
\affiliation{$^{}$Department of Physics, Pukyong National University, \\
Pusan 608-737, Korea\\
$^{a)}$Forecast Research Laboratory, Meteorological Research\\
Institute, KMA, Seoul 156-720, Korea\\
$^{b)}$Department of Pediatric Cardiology, Hanyang University,\\
Kuri 471-701, Korea \\
$^{c)}$Department of Physics, Kyushu University, \\
Fukuoka 812-81, Japan \\
}

\begin{abstract}

We study the decay process for the reaction-diffusion process of
three species on the small-world network. The decay process is
manipulated from the deterministic rate equation of three species
in the reaction-diffusion system. The particle density and the
global reaction rate on a two dimensional small-world network
adding new random links is discussed numerically, and the global
reaction rate before and after the crossover is also found by
means of the Monte Carlo simulation. The time-dependent global
reaction rate scales as a power law with the scaling exponent
$0.66$ at early time regime while it scales with $-0.50$ at long
time regime, in all four cases of the added probability
$p=0.2-0.8$. Especially, our result presented is
compared with the numerical calculation of regular networks.\\

\end{abstract}

\maketitle

\section { Introduction }

In the past decades, several papers have been devoted to the
theoretical and numerical investigation on the reactant
segregation phenomena in anomalous kinetics.$^{1-3}$ An important
contribution to this field of research has been given by
Ovchinnikov and Zeldovich,$^{4}$ who developed the segregation
phenomenon. It is well known that the binary reaction has
primarily been investigated on the process of ternary
reactions$^{5}$ and many phenomena in nature.$^{6-10}$ Until now,
the segregation of reactants has been investigated a
reaction-diffusion process based on assumption that the scaling
form for $A+B\rightarrow C$ has almost been verified by the
computer simulation and the experiment.$^{11-15}$ The decay
process for the reaction-diffusion system have been investigated
by Taitelbaum $et$ $al.$$^{16}$ who have precisely shown both the
global reaction rate and the reaction front increased as a
function of $t^{1/2}$ at very early times. At that time, Cornell
$et$ $al.$$^{17,18}$ have argued the diffusion-limited reaction
$nA+mB \rightarrow C$ for both homogeneous and inhomogeneous
conditions under initially separated reactants. They have
discussed that the global reaction rate decreases as $t^{-1/2}$ in
long time limit, independent of $n$ and $m$, and that the upper
critical dimension is $d_c =2$ for the reaction-diffusion process.

Recently, Zumofen $et$ $al.$$^{19}$ have studied mainly the
particle density and the pair correlation function on the
two-particles reaction process. Particularly, they found that the
particle density is distributed as $P(n)\sim n^{-1-\gamma}$,
$n>0$, $1<\gamma<2$, for L\'{e}vy walks and that the segregation
disappears in $d=3$ dimension for $\gamma<3/2$. Yen $et$
$al.$$^{20}$ have studied for the asymptotic early-time scaling in
the ternary reaction-diffusion process with initially separated
reactants. Moreover, Kim $et$ $al.$$^{21}$ argued the decay
process in a reaction-diffusion system with three species on the
regular square lattice. By a simple perturbation expansion, we
have analytically derived the particle density and the global
reaction rate before and after the crossover in the
reaction-diffusion system of $A + B + C \rightarrow 0 $.

For last few years, a growing interest has been concentrated on
small-world and scale-free network models,$^{22}$ and these models
have recently been widely studied in various applications of
methods from physics to other natural, social and applied
sciences. In fact, the phenomena of small-world and scale-free
networks are showed to be different from the dynamical behavior of
the regular lattice system. The degree distribution for scale-free
networks scales as a power law $p(k)\sim k^{-\gamma}$ for large k,
while it decays faster than exponentially for random networks. It
is, in present, of fundamental importance that the numerical and
analytical result of small-world and scale-free network models is
compared with that of regular network models.

Recently, the bimolecular chemical reaction in scale-free
networks$^{23}$ is in scale-free networks studied the generation
of the depletion zone and the segregation of the reactants. It was
found that the reaction-diffusion processes in scale-free networks
is different in their nature compared to regular lattice models,
due to the small diameter of networks and the existence of hubs.
Furthermore, Catanzaro $et$ $al.$$^{24}$ have analyzed that the
inverse particle density scales linearly as $1/{\rho(t)} \sim t$.
From this result, they have found that the inverse particle
density in uncorrelated scale-free network crosses over to a
linear behavior. Very recently, Gallos and Argyrakis$^{25}$ have
discussed the reaction-diffusion process of the two species on the
scale-free network between the the correlated and the uncorrelated
configuration models. They have especially revealed that two
models are identical when $\gamma=3.0$.

In this paper, the decay process for the reaction-diffusion
process of three species on the small-world network is studied. We
also consider the particle density and the global reaction rate on
the two-dimensional small-world lattices added by new random
links. In Section $2$, we discuss the reaction-diffusion process
of three species on small-world networks. We present some results
obtained by the numerical simulations and the concluding remark in
the final section.

\section {Reaction-diffusion system
of three species   }

In the reactions of $A+A\rightarrow 0$ and $A+B\rightarrow 0$
types on a $d$ dimensional regular lattice, the surviving particle
density scales in long time limit as
\begin{equation}
\frac{1}{\rho(t)}-\frac{1}{{\rho}_0} \sim t^{\alpha},
 \label{eq:a11}
\end{equation}
where ${\rho}_0$ is the initial particle density. It is well known
that the scaling exponent $\alpha=d/d_c$ for $d\leq d_c$ and $1$
for $d>d_c$, and that the critical dimension $d_c =2$ for $A+A$
and $4$ for $A+B$.

Next the $A+B\rightarrow 0$ process on a regular network$^(19)$ is
introduced for L\'{e}vy walks using the following
reaction-diffusion equation,
\begin{equation}
\frac{\partial{}}{\partial{t}} A(\vec{r},t)  =
D\hat{L}A(\vec{r},t) - \kappa A(\vec{r},t)B(\vec{r},t),
\label{eq:a1}
\end{equation}
\begin{equation}
\frac{\partial}{\partial t} B(\vec{r},t)  = D\hat{L}B(\vec{r},t) -
\kappa A(\vec{r},t)B(\vec{r},t), \label{eq:b1}
\end{equation}
where $A(\vec{r},t)$ and $B(\vec{r},t)$ are the particle
densities, $D$ a generalized diffusion coefficient, $\gamma$ a
reaction rate, and $\hat{L}$ the operator for the
L\'{e}vy-enhanced diffusion. The time-dependent particle densities
can be calculated as
\begin{equation}
A(t) = B(t) \sim t^{d/2\gamma}, \,\, \textrm{for} \, \,
\gamma>d/2,
\label{eq:bb1}
\end{equation}
%
where three marginal values are $\gamma=1$ for
$d=2$, $\gamma=3/2$ for $d=3$, and $\gamma=2$ for $d=4$.

From the decay process$^{20}$ of $A+2B$$\rightarrow$$C$, the
diffusion equations under the initial reactant segregation are as
follows:
\begin{equation}
\frac{\partial}{\partial t}A(x,t)  = D_a \nabla^2 A(x,t) - k
A(x,t)B^2 (x,t), \label{eq:c1}
\end{equation}
\begin{equation}
\frac{\partial}{\partial t}B(x,t)  = D_b \nabla^2 B(x,t) - k
A(x,t)B^2 (x,t), \label{eq:d1}
\end{equation}
where $A(x,t)$ and $B(x,t)$ are the particle densities, $k$ is the
microscopic reaction constant, and $D_a$ and $D_b$ are the
diffusion coefficients of reactants $A(x,t)$ and $B(x,t)$,
respectively. Then, from the lowest order of the perturbation
theory, the global reaction rates $R(t)$ on both the early time
and long time behaviors scaled as a power law $ t^{1/2}$ and $
t^{-1/2}$, respectively. Using the Monte carlo method, it was
found that the slopes are, respectively, $0.5$ and $-0.48$ before
and after the crossover.

Let us denote that $ A (x,t)$,  $B (x,t)$, and  $C (x,t)$ are the
particle densities for three-species $A$, $B$, and $C$ existing at
a position $x$ at time $t$. We assume that three species are
initially distributed separately on the axis $x$. Then, the
deterministic rate equation for $A (x,t)$ can be expressed in
terms of
\begin{equation}
\frac{\partial}{\partial t} A (x,t) =  D_A {\nabla}^2  A(x,t) - K
A (x,t)  B (x,t)  C (x,t),
\label{eq:e1}
\end{equation}
where  $D_A$ is the diffusion constant for one species $A$, and
$K$ is the reaction rate. The solution for $A(x,t)$, $B(x,t)$, and
$C(x,t)$ is obtained that
\begin{equation}
A(x,t)= B(x,t)=C(x,t) \sim  \phi(\frac{x}{{t}^{1/2}})
\label{eq:g7}
\end{equation}
in large time limit, where $\phi(x) = \frac{1}{\sqrt{\pi}}
\int^{x}_{-\infty} dxe^{-x^{2}}$. Similar to the result of the
previous work,$^{20}$ the time dependence of the global reaction
rates $R(t)$ was shown to behave as
\begin{equation}
R(t) \sim t^{1/2} \, \,   \textrm{and} \,\, {t}^{-1/2}
\label{eq:g78}
\end{equation}
in early time and long time limits.$^{21}$

\section { Numerical results and concluding remarks }

\begin{figure}[]
\includegraphics[angle=90,width=9.0cm]{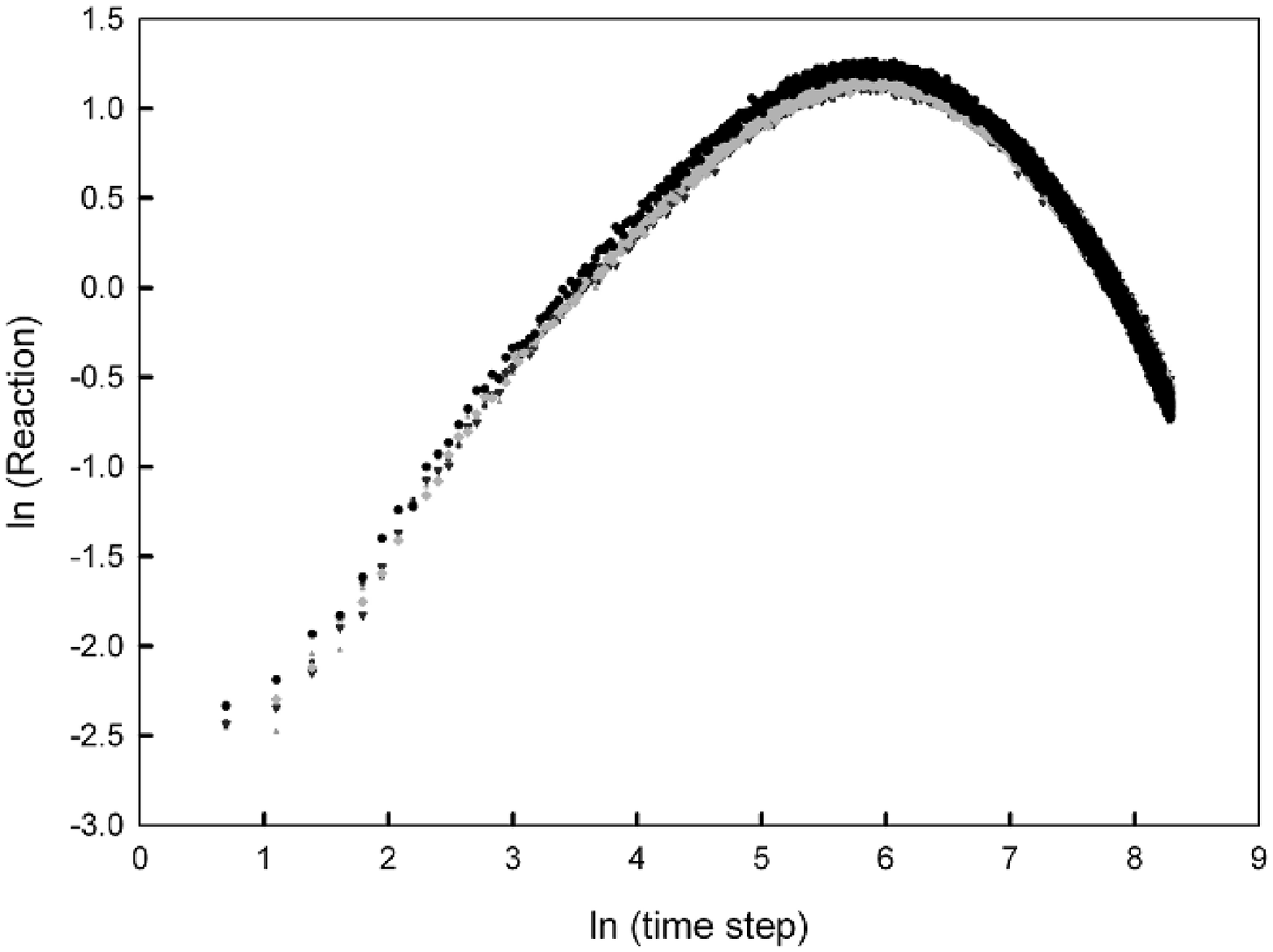}
\caption[0]{Plot of the global reaction rate $R(t)$ for the
reaction-diffusion process of $A+B+C\rightarrow 0$. Our numerical
simulation is performed on $2\times10^3$ configurations in
$200\times200$ square lattice that is added by new random links
for the added probability $p=0$ (black circles), $0.2$ (squares),
$0.4$ (circles), $0.6$ (down triangles), and $0.8$ (up
triangles).}
\end{figure}

In order to confirm numerically the global reaction rate, we
mainly use the Monte Carlo simulation method in reaction-diffusion
process of $A + B + C \rightarrow 0 $ on the small-world network
adding new random links. First of all, we assume that three
species reactants are distributed randomly on two dimensional
square lattice with the periodic boundary condition. After one
species reactant is chosen at random, the direction of its move is
chosen at random with equal probability to one of its linked
neighbor lattice points. When two species reactants meet each
other on the same lattice point, the intermediate process existing
concurrently the combined two species reactants can be formed. If
two species reactants meet the third reactant, these reactants
react and leave immediately on lattice point. If one species or
two species reactants cannot meet the other reactant, the
reactants diffuse randomly to one of its linked neighbor lattice
points. The diffusion constant in our case takes the same value
for each reactant, and the respective particle density of
$13.33(40/3)\%$ for $A$, $B$, and $C$ is distributed randomly on a
square lattice having $200 \times 200$ lattice points with the
boundary condition. After our simulation is performed on
$2\times10^3$ realizations and $K=1/1500$, we directly observed
the crossover for the global reaction rate from our simulation
result. It is numerically found that the scaling exponent for the
slope before and after the crossover is, respectively, $0.66$ and
$-0.54$ on the regular network $(p=0)$. As shown in Fig. $1$, the
time-dependent global reaction rate scales as a power law with the
scaling exponent $0.66$ at early time regime while it scales with
$-0.50$ at long time regime, in all four cases of the added
probability $p=0.2$, $0.4$, $0.6$, and $0.8$.

In conclusion, we have numerically estimated the global reaction
rate before and after the crossover in reaction-diffusion system
of $A + B + C \rightarrow 0 $ on square lattice added by new
random links. It is really found from our simulation result that
the scaling exponent of the global reaction rate on small-world
network added by new random links is the similar to that of the
regular network at early time regime. At long time regime, the
decay process on small-world network proceeds slightly faster than
in the case of the regular network.

In future, our work is in progress to extend the correlated and
the uncorrelated configuration models. We also will attempt to
investigate small-world and scale-free networks in several
scientific fields for three species reaction-diffusion process.

\begin{acknowledgements}
This work was supported by Pukyong National University Research
Fund in 2005.
\end{acknowledgements}

%
%
%
%

%


\begin{thebibliography}{}
%
\bibitem{Tou1} D. Toussaint and F. Wilczek, J. Chem. Phys. {\bf 78}, 2642 (1983).
\bibitem{Blu2} A. Blumen, J. Klafter, and G. Zumofen, $in$ $Optical$ $Spectroscopy$
                $of$ $Glasses$, edited by I. Zschokke(Reidel, Dordrecht, Holland, 1986)
\bibitem{Mea3} P. Meakin and H. E. Stanley, J. Phys. {\bf A 17}, L173 (1984).
\bibitem{Ovc4}  A. A. Ovchinnikov and Y. B. Zeldovich, Chem. Phys. {\bf 28}, 215 (1978).
\bibitem{Yen5} A. Yen and R. Kopelman, Phys. Rev. {\bf E 56}, 3694
(1997).
\bibitem{Hen6} H.K. Henisch, $Crysatal$ $in$ $Gels$ $and$ $Liesegang$ $rings$
                ( Cambridge University Press, Cambridge, England, 1988).
\bibitem{Avn7} D. Avnir and M. Kagan, Nature {\bf 307}, 717 (1984).
\bibitem{Dee8} G. T. Dee: Phys. Rev. Lett. {\bf 57}, 275 (1990).
\bibitem{Hei9} B. Heidel, C. M. Knobler, R Hilfer, and R. Bruinsma, Phys. Rev. Lett.
 {\bf 60}, 2492 (1988).
\bibitem{Mue10} K. F. Mueller, Science {\bf 225}, 1021 (1986).
\bibitem{Gal11} L. Galfi and Z. Racz, Phys. Rev. A {\bf 38}(1988) 315.
\bibitem{Koo12} Y.E. L. Koo, L. Li, and R. Kopelman, Mol. Cryst. Liq. Cryst. {\bf 183}, 187 (1990).
\bibitem{Kop13} Y.E. Koo and R. Kopelman, J. Stat. Phys. {\bf 65}, 893 (1991).
\bibitem{Yen14} A. Yen, Y.E. L. Koo, and R. Kopelman, Phys. Rev. {\bf E 54}, 2447 (1996).
\bibitem{Jia15} Z. Jiang and C. Ebner, Phys. Rev. {\bf A 42}, 7483
(1990).
\bibitem{Tai16} H. Taitelbaum, S. Halvin, J. E. Kiefer, B. Trus, and G. Weiss, J. Stat. Phys. {\bf 65}, 573 (1991).
\bibitem{Cor17} S. Cornell, M. Droz, and B. Chopard, Phys. Rev. {\bf A 44}, 4826 (1991).
\bibitem{Cor18} S. Cornell, M. Droz, and B. Chopard, Physica {\bf A 188}, 322 (1992).
\bibitem{Zum19} G. Zumofen, J. Klafter and M. F. Shlesinger, Phys. Rev. Lett. {\bf 77}, 2830 (1996).
\bibitem{Yen24}  A. Yen, Z.-Y. Shi and R. Kopelman, Phys. Rev. {\bf E 57}, 2438 (1998).
\bibitem{Kim25} Kyungsik Kim, K. H. Chang and Y. S. Kong, J. Phys. Soc. Jpn. {\bf 68}, 1450 (1999).
\bibitem{Alb20} R. Albert and A.-L. Barabasi, Rev. Mod. Phys. {\bf 74}, 47 (2002);
S. N. Dorogovtsev  and J. F. F. Mendes, Adv. Phys. {\bf 51}, 1079
(2002); M. E. J. Newman, SIAM Review {\bf 45}, 167 (2003).
\bibitem{Gal21} L. K. Gallos and P. Argyrakis, Phys. Rev. Lett. {\bf 92}, 138301 (2004).
\bibitem{Yen22} M. Catanzaro, M. Bogu\~{n}\'{a} and R. Pastor-Satorras, cond-mat/0407447.
\bibitem{Yen23}  L. K. Gallos and R. Kopelman, cond-mat/0503234.
%
\end{thebibliography}
\end{document}